\documentclass[prb,aps,twocolumn,showpacs]{revtex4-1}%
\usepackage{amsmath}
\usepackage{amsfonts}
\usepackage{amssymb}
\usepackage{graphicx}%
\usepackage{bm}
\usepackage{color}
\begin{document}
\title{Meissner response of superconductors with inhomogeneous penetration depths}
\author{V. G. Kogan$^1$ and J. R. Kirtley$^2$ }
%\email{kogan@ameslab.gov}
 \affiliation{$^1$Ames Laboratory DOE, Ames, IA 50011}
 \affiliation{$^2$Geballe Laboratory for Advanced Materials and  Department of Applied Physics  Stanford University, Stanford, California 94305-4045, USA}

\date{\today, started Dec.1, 2010  }
% \today 
\begin{abstract}
We discuss the Meissner response to a known field source of superconductors having inhomogeneities in their penetration depth. We simplify the general problem by assuming that the perturbations of the fields by the penetration depth inhomogeneities are small. We present expressions for inhomogeneities in several geometries, but concentrate for comparison with experiment on planar defects, perpendicular to the sample surfaces, with superfluid densities different from the rest of the samples.  These calculations are relevant for magnetic microscopies,  such as Scanning Superconducting Quantum Interference Device (SQUID)  and  Magnetic Force Microscope,  which image the local diamagnetic susceptibility of a sample.
%an Interaction of a vortex with the plane defect having enhanced or suppressed superfluid density is discussed.
\end{abstract}

\maketitle

 \section{Introduction}
    
  Recently Kalisky {\it et al.} reported the observation of ``stripes" of enhanced local diamagnetic susceptibility in scanning SQUID microscope (SSM) images of underdoped Ba(Fe$_{1-x}$Co$_x$)$_2$As$_2.$\cite{kalisky2010} They interpreted these stripes as being due to enhanced superfluid density (reduced London penetration depth) along twin boundaries. Kirtley {\it et al.}  modeled these experiments by solving London's and Maxwell's equations using finite element methods in an appropriate geometry.\cite{kirtley2010} These calculations were in   agreement with the experimental lineshapes, and provided estimates for the increase in the superfluid  density on the twin planes. However,  it was difficult to model geometries with regions of enhanced superfluid density as narrow as seemed physically likely, so that extrapolations to narrow widths from wider ones were necessary. 
 
The geometry relevant to this problem, that of a slab imbedded perpendicular to the sample surface in a bulk half-space superconductor, is difficult to treat analytically.
Here we make the problem tractable by  (1) assuming that the width of the region with reduced penetration depth is small relative to  other lengths in the problem, and (2) treating the problem to first order in a perturbation expansion. The first assumption is most likely valid for the case of SSM, since in this case the experimentally observed stripes in susceptibility are resolution limited.\cite{kalisky2010} 
Treating the problem to first order in a perturbation expansion seems reasonable, since at least at low temperatures the stripes in susceptibility observed using SSM are much smaller than the susceptibility itself.

 \subsection{ SSM technique}
 
 Although the method developed here for the evaluation of the Meissner response  of superconductors with inhomogeneities is general, we will use as a concrete example scanning SQUID susceptometry, \cite{gardner2001} which employs a sensor with two concentric, co-planar loops: one loop carries a small current $I$ that is a source of a weak magnetic field, and the other loop couples the response magnetic flux into the sensor SQUID. This is an elaboration of the common SQUID magnetometry, in which a SQUID senses the intrinsic magnetic fields without a source coil. Our results are relevant for magnetic force microscopy (MFM) as well. However, applying our approach to MFM requires modeling of magnetic tips with complex geometry and is outside of the scope of this paper. 
 
Twin and grain boundaries in superconductors may have enhanced as well as suppressed superfluid density.    
We will use below a generic term ``defect". The theory developed here applies for both enhancement and suppression, provided that the deviation of the superfluid density (or of the London penetration depth) at the defect from the bulk value is small.  

\subsection{Method}

Let us consider a magnetic field source with known field distribution   ${\bm
h}^s $ in the absence of a superconductor. The source is placed above 
  the superconducting half-space  
$ z<0$.   The total field in the empty half-space $z>0$ can be written as 
\begin{equation}
{\bm h} = {\bm h}^s + {\bm h}^r \,,
\end{equation}
 where ${\bm h}^r$ is the response field, which satisfies  ${\rm div} {\,\bm 
h}^r= {\rm curl} {\,\bm h}^r=0$ in vacuum outside the superconductor.  One can look for this field as
$\bm \nabla\varphi^r$, with the scalar potential $\varphi^r$ obeying the Laplace equation
and the boundary condition that it approaches zero far from the surface. The general form of such
a potential is
\begin{equation}
\varphi^r ({\bm r},z)=\int\frac{d^2{\bm k}}{(2\pi)^2}\,\varphi^r ({\bm k})\,
e^{i{\bm k}\cdot{\bm r}-kz}\,.
\label{Laplace}
\end{equation}
  Here, ${\bm r}=(x,y)$, ${\bm k}=(k_x,k_y)$, and $z$ is directed normal to the superconducting flat surface at $z=0$; $\varphi^r ({\bm k})e^{-kz}$ is the  two-dimensional (2D) Fourier
transform with respect to variables $x,y$ at any fixed  $z>0$. The potential
(\ref{Laplace}) is defined only in the upper half-space; hence, %**
 the problem of uniqueness that is in general associated with the description of the static 
magnetic field by a scalar potential does not arise. 

Thus, to know the outside field distribution it suffices to find the 2D Fourier transform $\varphi^r ({\bm k})$. In principle, this can be done by solving the internal London problem and 
by utilizing the boundary conditions of field continuity at the interface $z=0$. The formal difficulty to overcome  is to calculate the 2D Fourier transform of internal fields for non-uniform superconductors. Below, we show how this can be done in a few cases relevant for SSM and MFM. 

%%%%%%%%%%%    
 \subsection{Uniform and isotropic half-space}
 %%%%%%%%%
 
To demonstrate the method, we start with the simple situation of a uniform and isotropic half space, for which the  London equation   is $ {\bm h}_0-\lambda_0^2 \bm\nabla^2{\bm h}_0=0 $, with $\lambda_0$ being the London penetration depth.  
The 2D Fourier transform then reads:
\begin{eqnarray}
p^2{\bm h}_0({\bm k},z)-{\bm h}_0^{\prime\prime}({\bm k},z)=0 \,,\quad p^2= \lambda_0^{-2}+k^2\,,
\label{2DFT}
\end{eqnarray}
where the prime denotes $\partial/\partial z$. The solution that vanishes at $z\to -\infty$ is 
\begin{eqnarray}
 {\bm h}_0({\bm k},z)= {\bm H}({\bm k} )\,e^{pz}
 \label{solution}
\end{eqnarray}
with  ${\bm H}$  independent of $z$. ${\bm h_0}$ should satisfy ${\rm div}{\,\bm h}_0=0$, which yields in Fourier space:
\begin{eqnarray}
i(k_x H_x+k_y H_y)+pH_z=0\,.
 \label{divH=0}
\end{eqnarray}
The requirement of field continuity at $z=0$ gives:
\begin{eqnarray}
 H_x&=&ik_x (\varphi^s+\varphi^r_0) \,, \label{Hx}\\
 H_y&=&ik_y (\varphi^s+\varphi^r_0) \,, \label{Hy}\\
 H_z&=&k  (\varphi^s-\varphi^r_0 ) \,. \label{Hz}
\end{eqnarray}
We took into account here that for the source of the magnetic field placed at $z=z_0$, the potential under it, in particular at $z=0$, is given by 
\begin{equation}
\varphi^s({\bm r},z)=\int\frac{d^2{\bm k}}{(2\pi)^2}\,\varphi^s ({\bm k})\,
e^{i{\bm k}\cdot{\bm r}+k(z-z_0)}\,,
\label{Laplace1}
\end{equation}
so that $h_z^s({\bm k})=+k\varphi^s ({\bm k})$.

Multiplying Eq.\,(\ref{Hx}) by $ik_x$, (\ref{Hy}) by $ik_y$, and (\ref{Hz}) by $p$, adding them up, and using Eq.\,(\ref{divH=0}) yields:
\begin{eqnarray}
 \varphi^r_0({\bm k} ) = \frac{p-k}{p+k}\,  \varphi^s({\bm k} )\,,\qquad\qquad  \label{phi_r0}\\
 H_{x,y}=\frac{2ik_{x,y}p}{p+k} \varphi^s\,,\qquad H_z=\frac{2 k^2}{p+k}\varphi^s\,.
 \label{result0}
\end{eqnarray}
Thus, the response fields outside and inside are expressed in terms of the unperturbed source field $\varphi^s$.
 This result has been obtained in Ref.\,\onlinecite{Meissner} as a particular case of cumbersome anisotropic formulas; here it follows directly from the isotropic London equations.

%%%%%%%%    
 \section{Planar defect}
 %%%%%%%%%
 
For the general case of an inhomogeneous penetration depth $ \lambda(\bm r)$,  the magnetic field within the superconductor obeys the London equation in the form:
\begin{eqnarray}
 {\bm h}+\frac{4\pi}{c} {\rm curl}(   \lambda^2 {\bm j}) 
 ={\bm h}+ \frac{4\pi}{c}  \lambda^2 {\rm curl}{\,\bm j}+  \bm\nabla \lambda^2\times\frac{4\pi}{c}{\bm  j} =0.\nonumber\\
\label{L1}
\end{eqnarray}
This equation is the minimum condition for the London energy functional
\begin{equation}
\varepsilon_L =\int  \frac{dV}{8\pi } \left[h^2+  \lambda ^2 ({\rm curl}{\,\bm h})^2\right],
\label{Lond-energy}
\end{equation}
which holds for inhomogeneous $\lambda$. 

For a planar defect at $x=0$, we  model the penetration depth by 
\begin{equation} 
\lambda^2(x) = \lambda^2_0 -  \beta^3 \delta(x)\,,
\label{lam(x)}
\end{equation}
where a positive $\beta$ with the dimension of length  is related to a superfluid density enhancement, whereas $\beta<0$ corresponds to a superfluid density suppression. Physically, the superfluid density  at the planar defect may extend to distances on the order of the coherence length $\xi$ into the bulk. However, within the London approach for materials with $\xi\ll\lambda$ the representation (\ref{lam(x)}) is justified. The advantage of  Eq.\,(\ref{lam(x)}) is that it allows one to do the 2D Fourier transform of the London equation for which  analytic expressions for all transformed quantities on the whole $x,y$ plane are needed. 

An alternative way to address the problem could be to consider the defect as a   layer of a finite thickness with the penetration depth different from $\lambda_0$ of the surrounding material, to look for solutions of the London equations in each part separately and to match them with certain boundary conditions. These real space solutions should then be matched with the real space field distribution  in the outer space to calculate the response field. This approach, however, is  more cumbersome and certainly less tractable and transparent as compared to the method utilizing the 2D Fourier transform employed here.   

With $\lambda(x)$ of Eq.\,(\ref{lam(x)}), the London equation (\ref{L1}) takes the form
\begin{eqnarray}
 {\bm h}- \lambda_0^2 {\rm \nabla^2}{\bm h} =\beta^3\delta^\prime(x)\hat {\bm x}\times {\rm curl} {\,\bm h}-\beta^3\delta(x) {\rm \nabla^2}{\bm h}.\qquad
\label{L3D}
\end{eqnarray}
The idea of the following manipulation is based on the physical assumption that  the influence of the defect on the field distribution is weak, $\beta\ll \lambda_0$, and one can use  a perturbation argument for its evaluation. 
 Fits of the present theory to the experiments of Kalisky {\it et al.} (Fig. \ref{fig:3}) require values of $\beta \sim \lambda_0$. However, comparison of finite element modeling of the same problem (Fig. \ref{fig:2}) are in reasonable agreement with the present theory, even for $\beta \sim \lambda_0$. This justifies keeping only the first order in perturbation theory, resulting in a considerable simplification of the problem.

Having this in mind, we look  for the field inside as ${\bm h}={\bm h}_0+{\bm   h}_b$, where the unperturbed field satisfies $ {\bm h}_0- \lambda_0^2 {\rm \nabla^2}{\bm h}_0 =0$ in the absence of the defect plane, whereas ${\bm  h}_b$ is a perturbation due to the boundary. We then obtain in the first order:
\begin{eqnarray}
 {\bm h}_b- \lambda_0^2 {\rm \nabla^2}{\bm h}_b =\beta^3\delta^\prime(x)\hat {\bm x}\times {\rm curl} {\bm h}_0  -\beta^3\delta(x) {\rm \nabla^2}{\bm h}_0,\qquad
\label{L3Da}
\end{eqnarray}
where ${\bm h}_0$ has been calculated in the preceding  section. 

One now calculates the 2D Fourier transform (FT) of the left-hand side (LHS):
\begin{eqnarray}
 FT( {\bm h}_b- \lambda_0^2 {\rm \nabla^2}{\bm h}_b)  = (1+\lambda_0^2k^2){\bm h}_b({\bm k},z) - \lambda_0^2 {\bm h}_b^{\prime\prime}({\bm k},z). \qquad
\label{FTa}
\end{eqnarray}

Calculating  the 2D FT of   the RHS of  Eq.\,(\ref{L3Da}), one can use easily verifiable identities, see Appendix A:
\begin{eqnarray}
& & FT[\delta (x)f(\bm r) ]=
  \int_{-\infty}^\infty \frac{dq_x}{2\pi} \,f(q_x,k_y)\,,\label{id1} \\
& & FT[\delta^\prime(x)  f(\bm r) ] = i  \int_{-\infty}^\infty \frac{dq_x}{2\pi} \, (k_x-q_x)f(q_x,k_y)\,.\quad\label{id2}
\end{eqnarray}
We obtain after straightforward algebra:
\begin{eqnarray}
p^2{\bm h}_b({\bm k},z) - {\bm h}_b^{\prime\prime}({\bm k},z)=\frac{
\beta^3}{ \lambda_0^2}  \int_{-\infty}^\infty \frac{dq_x}{2\pi}   {\bm A}  \,, 
\label{eq40}
\end{eqnarray}
where out of the three components of the vector ${\bm A}$ we will need only one:
\begin{eqnarray}
A_{ z} =  ( ih_{0x}^\prime +q_x h_{0z} )(k_x-q_x) - h_{0z}^{\prime\prime} + Q^2h_{0z} \,,
\label{Az}
\end{eqnarray}
where ${\bf Q}=(q_x,k_y)$.
 The field ${\bm h}_0({\bm Q},z)$ satisfies Eq.\,(\ref{2DFT}) in which one should replace ${\bm k}\to {\bm Q}$ and $p\to K $: 
\begin{eqnarray}
{\bm h}_0^{\prime\prime}({\bm Q},z)=K^2{\bm h}_0({\bm Q},z) \,,\quad K=\sqrt{\lambda_0^{-2}+Q^2}\,.
 \label{K}
\end{eqnarray}
Hence,
\begin{eqnarray}
{\bm h}_0 ({\bm Q},z) ={\bm H} ({\bm Q}) e^{Kz}\,
 \label{h0}
\end{eqnarray}
with ${\bm H}$ given in Eq.\,(\ref{result0}):
\begin{eqnarray}
 H_{x,y}=\frac{2iQ_{x,y}K}{K+Q} \varphi^s({\bm Q})\,,\quad H_z=\frac{2 Q^2}{K+Q}\varphi^s({\bm Q})\,.\quad
 \label{H's}
\end{eqnarray}
Substituting  these ${\bm H} ({\bm Q})$ in  Eq.\,(\ref{Az}) we obtain:
\begin{eqnarray}
% &&A_{x,y}=-\frac{2iQ_{x,y}K}{\lambda_0^2(K+Q)} \varphi^s({\bm Q})\,,\nonumber\\ 
A_z=- 2 (K-Q) ( \bm Q\cdot\bm k) e^{Kz}\varphi^s({\bm Q})=A_{0z}e^{Kz}\,.\qquad
 \label{A's}
\end{eqnarray}
 
We now write Eq.\,(\ref{eq40}) for the field perturbation in a compact form:
\begin{eqnarray}
   {\bm h}_b^{\prime\prime}({\bm k},z)-p^2{\bm h}_b({\bm k},z)=
{\bm D}({\bm k},z) ,\nonumber\\
{\bm D}({\bm k},z)=-\frac{\beta^3}{\lambda_0^2}  \int_{-\infty}^\infty \frac{dq_x}{2\pi} e^{Kz}  {\bm A}_0\,.
\label{dif}
\end{eqnarray}
This is a second order linear differential equation for ${\bm h}_b({\bm k},z)$ with respect to the variable $z$. The  solution vanishing at $z\to -\infty$  is 
\begin{eqnarray}
   {\bm h}_b({\bm k},z)= {\bm C}e^{pz}
- \frac{ \beta^3}{\lambda_0^2}  \int_{-\infty}^\infty \frac{dq_x}{2\pi} \frac{e^{Kz}}{K^2-p^2} \, {\bm A}_0 \qquad
\label{solution}
\end{eqnarray}
(see Appendix B).
The arbitrary vector ${\bm C}=(C_x,C_y,C_z)$   is to be determined from the boundary conditions. 

In fact, the constants $C_i$ are not independent because div$ {\bm h}_b=0$. 
In particular, at $z=0$ this gives
\begin{eqnarray}
 i{\bm k}\cdot{\bm C}+pC_z=
  \frac{ \beta^3}{\lambda_0^2}  \int_{-\infty}^\infty \frac{dq_x}{2\pi} \frac{(i{\bm k}\cdot{\bm A}_0+KA_{0z})}{K^2-p^2} \,  .\qquad
\label{div=0}
\end{eqnarray}
 
Now, we can formulate the boundary conditions of field continuity at $z=0$:
\begin{eqnarray}
i k_x(\varphi^s+\varphi^r) =h_{0x}+C_x-\frac{ \beta^3}{\lambda_0^2}  \int_{-\infty}^\infty \frac{dq_x}{2\pi} \frac{A_{0x}}{q_x^2-k_x^2} \,,\qquad\label{x}\\
  ik_y (\varphi^s+\varphi^r) =h_{0y}+C_y-\frac{ \beta^3}{\lambda_0^2}  \int_{-\infty}^\infty \frac{dq_x}{2\pi} \frac{A_{0y}}{q_x^2-k_x^2} \,,\qquad\label{y}\\
  -k  (\varphi^r-\varphi^s) =h_{0z}+C_z-\frac{ \beta^3}{\lambda_0^2}  \int_{-\infty}^\infty \frac{dq_x}{2\pi} \frac{A_{0z}}{q_x^2-k_x^2} \,.\qquad\label{z}.
 \label{bc}
\end{eqnarray}
Multiply the first equation by $ik_x$, the second by $ik_y$, and the third by $p$ and add them up. The terms with ${\bm h}_0$ add to zero because div${\,\bm h}_0=0$. Utilizing Eq.\,(\ref{div=0}) we obtain for the defect contribution to the outside magnetic potential:
\begin{eqnarray}
&&\psi(\bm k)=\varphi^r(\bm k)-\varphi^r_0(\bm k)=\varphi^r(\bm k)-\frac{p-k}{p+k} \varphi^s(\bm k) \qquad \nonumber\\
&&=  \frac{2 \beta^3}{\lambda_0^2k(k+p)}  \int_{-\infty}^\infty \frac{dq_x}{2\pi} \frac{(K-Q) {\bm k}\cdot{\bm Q} }{ K+p } \varphi^s(\bm Q)\, .\qquad
\label{otvet}
\end{eqnarray}
 
%%%%%%%%%%%
\section{Application to SQUID susceptometry}
%%%%%%%%
 
\begin{figure}[t]
        \begin{center}
                \includegraphics[width=9cm]{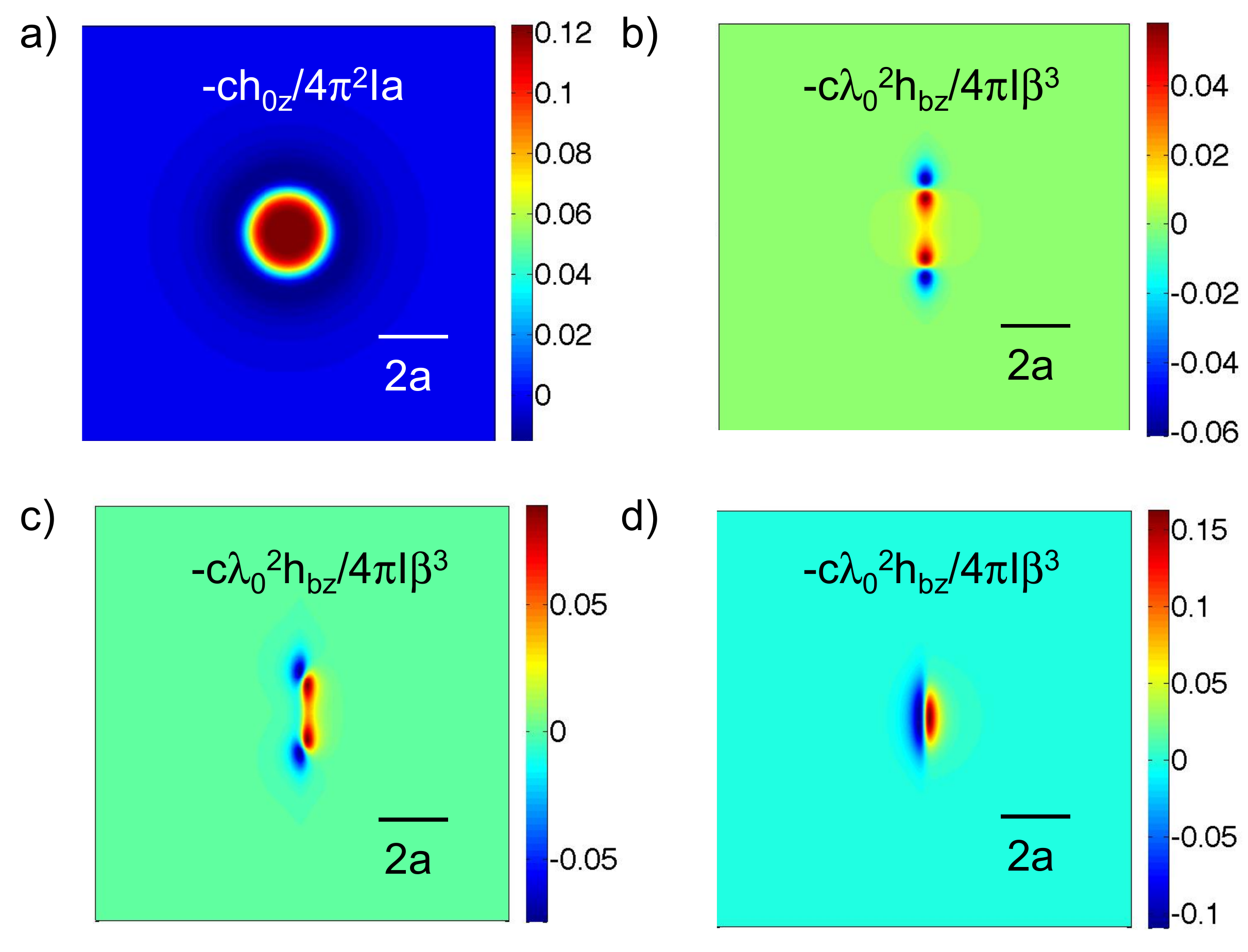}
        \end{center}
        \caption{(Color online) Calculated SQUID susceptometer response fields: a) $-ch_{0z}(x,y)/4\pi^2Ia$, where $h_{0z}$ is the $z$-component of the field at the SQUID plane, $I$ is the current through the field coil of a radius $a$, in the absence of a planar defect.  b) $-c\lambda_0^2 h_{bz}/4\pi I \beta^3$, where $h_{bz}$ is the $z$-component of the field due to the planar defect, $\lambda_0$ is the London penetration depth of the bulk superconductor, and $\beta$ determines the size of the change in penetration depth at the planar defect, Eq.\,(\ref{lam(x)}), for $x_0/a=0$, c) $x_0/a=1/2$, and d) $x_0/a=1$. Here $z_0/a=0.17$ and $\lambda/a=0.05$.}
        \label{fig:stripe_susc_images}
\end{figure}
 
The potential of a circular current source of the SQUID susceptometer is given by its 2D Fourier transform:\cite{Meissner}
\begin{equation}
\varphi^s(\bm k)=\frac{4\pi^2 I a}{ck} e^{-k z_0}J_1(k a)e^{-i k_x x_0},
\label{eq:source_susc}
\end{equation}
where $I$ is the current through the field coil of radius $a$,   $(x_0,0,z_0)$ are the coordinates of the coil center, and $z_0$ is the height of the coil above the sample surface. 
 
%%%%%%%%%%
 \subsection{Uniform sample}
%%%%%%%%%%
 
The potential of the response field is given in  Eq.\,(\ref{phi_r0}), so that  the 2D FT of the response field $h_{0z}(\bm k)=-k \varphi_0^r(\bm k)$ for  a   superconducting half-space free of defects is given by
\begin{equation}
h_{0z}( {\bm k},z_0) = -\frac{4\pi^2 I a}{c} \,  \frac{p-k}{p+k}   e^{- k z_0}  J_1(k a)\,;
\label{eq:response_susc}
\end{equation}
here we have set $x_0=0$ since all positions $x_0$ are equivalent in this case. 
This gives the distribution of the $z$ component of the field in the SQUID plane:
\begin{equation}
h_{0z}( {\bm r},z_0) = -\frac{I a}{c} \int\frac{d\bm k(p-k)}{p+k}   e^{i{\bm k}{\bm r}- 2k z_0  } J_1(k a). 
\label{eq:response_real}
\end{equation}
This distribution is shown in Fig.\,\ref{fig:stripe_susc_images}a for the parameters indicated in the caption. 
Integrating this over the SQUID loop area of radius $r$, we obtain the flux of the response field:
\begin{equation}
\Phi_0^r = -\frac{\pi I ar}{c } \int_0^\infty dk\,\frac{ p-k }{p+k}   e^{ - 2k z_0  } J_1(k a)J_1(k r)\,; 
\label{eq:response_real}
\end{equation}
where  $p^2=\lambda_0^{-2}+k^2$. 

%%%%%%%%%%
 \subsection{Planar defect}
%%%%%%%%%%

\begin{figure}[h]
        \begin{center}
                \includegraphics[width=7cm]{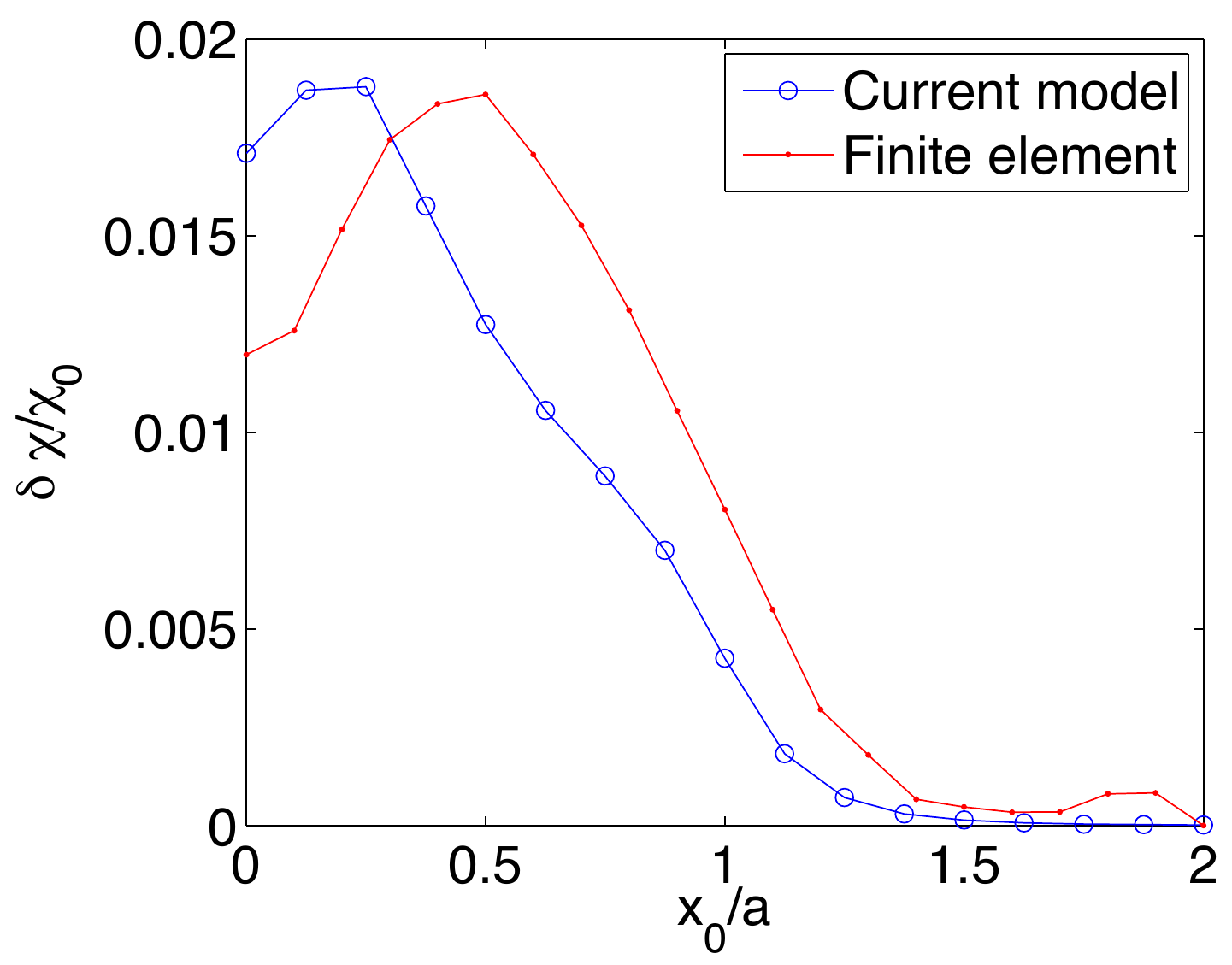}
        \end{center}
        \caption{(Color online) Calculated normalized SQUID susceptibility $\delta \chi/\chi_0$, where $\delta \chi$ is the change in the SQUID susceptibility due to the planar defect, and $\chi_0$ is the SQUID susceptibility in the absence of a planar defect, for $y_0=0$, with $r=0.25 a$, $z_0=0.17a$: The curve labelled ``Current model" (open symbols) evaluates Eq.'s (\ref{eq:response_susc}) and (\ref{eq:dresponse_susc}) with $\beta^3/\lambda_0^2 a =0.15$. The curve labelled ``Finite element" (closed symbols) is numerical modeling of a stripe with finite width $w/a=0.2$ with $\lambda_0/a=0.2$ and $\lambda_b/a=0.1$. }
        \label{fig:2}
\end{figure}
 
The Fourier transform of the $z$-component of the response field due to the planar defect is given by $\delta h_z( {\bm k})=-k\psi( {\bm k})$ with $\psi$  given in Eq.\,(\ref{otvet}) and $\varphi^s (\bm Q)$ obtained from Eq.\,(\ref{eq:source_susc}) with $\bm k$ replaced by $\bm Q$:
 \begin{eqnarray}
 &-&\frac{c\lambda_0^2} {4\pi I a \beta^3  }h_{bz}( {\bm k},z_0)  \label{eq:dresponse_susc}\\
&=&  \int_{-\infty}^{\infty} dq_x \frac{(K-Q) ( {\bm k}\cdot{\bm Q})J_1(Qa)}{(k+p) (K+p)Q}e^{-(Q+k)z_0 -iq_xx_0},\qquad \nonumber\\
  &&\bm Q=(q_x,k_y)\, ,\qquad p=\sqrt{k^2+1/\lambda_0^2}\,.\nonumber
\end{eqnarray}
 Here the integration over $q_x$ is done numerically for each $\bm k$ and the  results   
 are Fourier transformed to obtain the magnetic fields as a function of position in real space. Selected results for the fields are shown in Figure \ref{fig:stripe_susc_images}.  
Fig. \ref{fig:stripe_susc_images}a shows the response field $h_0(\bm r,z=0)$, for a bulk superconductor in the absence of a planar defect. Negative response fields (colored red) correspond to diamagnetic shielding. Fig.\,\ref{fig:stripe_susc_images}b-d display the {\it change} in the response field, $  h_{bz}(\bm r,z_0)$, due to a planar defect at various spacings $x_0$ between the center of the field coil and the defect position. 
 
Next,  $  h_{bz}(\bm r,z_0)$ is integrated numerically over the SQUID loop  of a radius $r$ centered at $(x_0, 0)$ to obtain the change in magnetic flux $\Phi_b$. The integration can also be done analytically, see Appendix C:
\begin{eqnarray}
\Phi_b = \frac{r}{2\pi}\int \frac{d{\bm k}}{k}h_{bz}({\bm k},z_0)   J_1(kr) e^{ik_xx_0}.
\label{Phi_S}
\end{eqnarray}
  
SQUID susceptibilities are defined as $\chi=\Phi_b/I\Phi_0$, where 
 $\Phi_0=h/2e$ is the superconducting flux quantum. The curve labelled ``Current model" in 
Fig.\,\ref{fig:2}  shows the change in susceptibility $\delta\chi$ due to a planar defect at $x=0$ divided by the susceptibility $\chi_0$ in the absence of a defect as a function of the position $x_0$ of the SQUID sensor, with fixed $y_0=0$, $z_0/a=0.17$, $r/a=0.25$, $\beta/a=0.18$ and $\lambda_0=0.025$. The parameters $\lambda_0$ and $\beta$ were chosen for convenience of comparison with finite element modeling to be discussed in Section \ref{sec:finite_element}.
  
\begin{figure}[htb ]
        \begin{center}
                \includegraphics[width=8cm]{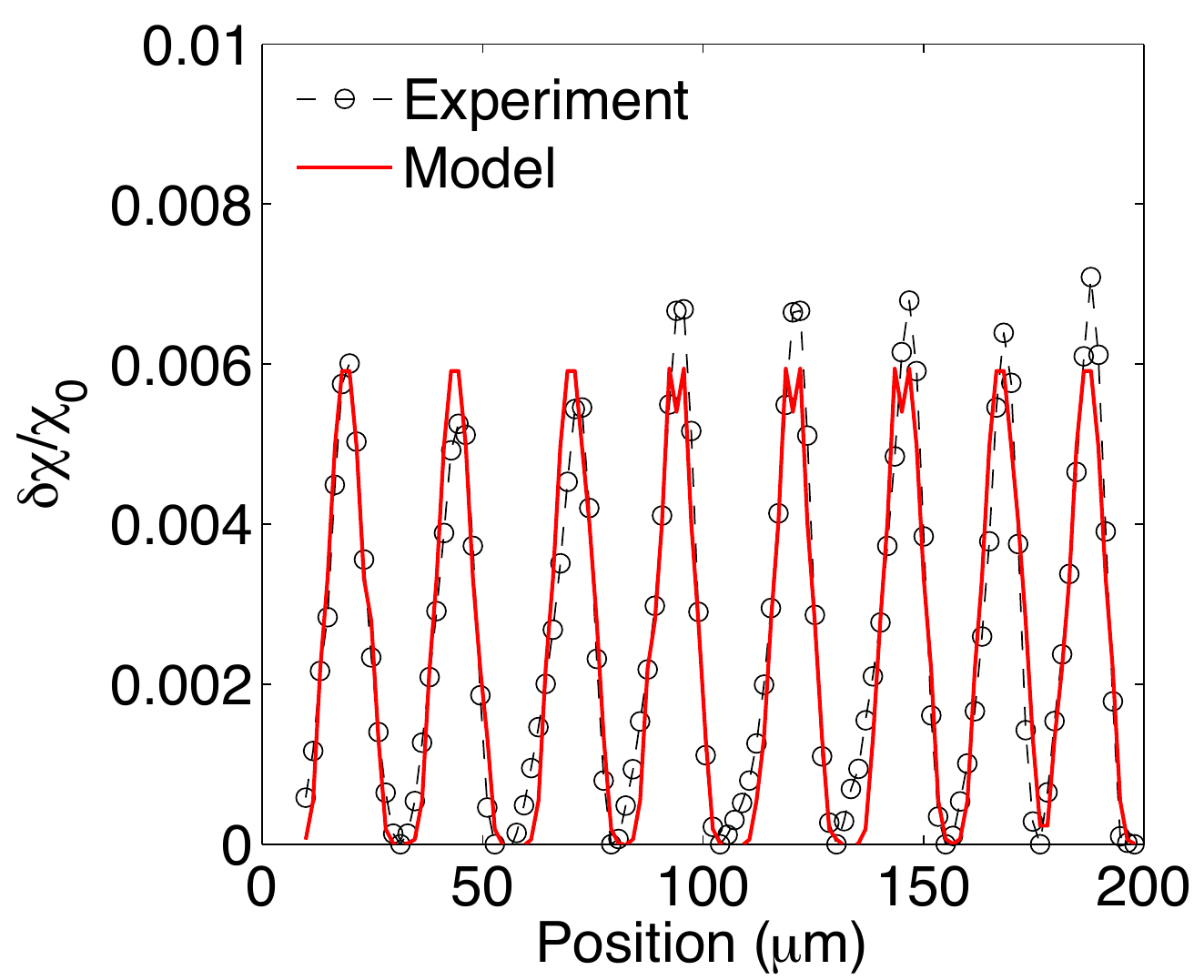}
        \end{center}
        \caption{(Color online) Fit of current model to experimental data of  Ref.\,\onlinecite{kalisky2010}. The fitting parameters are the positions in $x$ of the 8 peaks in susceptibility, an overall vertical shift, the radius of the field coil $a$, and a vertical scaling factor $\beta^3/\lambda_0^2 a$. The fixed parameters are $z_0/a=0.17$ and $r/a=0.25$, where $r$ is the radius of the SQUID pickup loop. The best fit values are $a = 7.1(+2-1.4)\,\mu$m and $\beta^3/\lambda_0^2 a = 0.048(+0.010-0.011)$, using a doubling of the best fit chi-square value as the criterion for determining the uncertainty in the fit parameters.}
          \label{fig:3}
\end{figure}
 
Figure \ref{fig:3} displays   the   predicted $\delta\chi/\chi_0$ versus $x_0$ along with   the  
data of Kalisky {\it et al.} taken on a twinned crystal of Ba(Fe$_{1-x}$Co$_x$)$_2$As$_2$. \cite{kalisky2010}  In this case, the fitting parameters were the positions of the twin boundaries, an overall scaling factor (corresponding to adjusting $\beta^3/\lambda_0^2 a$), the field coil radius $a$, and a vertical shift of the data. The fixed parameters were $z_0/a=0.17$ and $\lambda_0/a=0.05$.  The agreement between experiment and theory is reasonable.  The double maxima structure predicted by the theory is not observed in experiment, although this may be due to an insufficient signal-to-noise ratio. Also, we model the current ring and SQUID loop by linear circles whereas both of them have a width on the order of microns, making it difficult to resolve the structure of $\chi(x_0)$ on this length scale. The data are best fit by $0.037 < \beta^3/\lambda_0^2 a < 0.058$ and  5.7\,$\mu$m $< a <$ 9.1\,$\mu$m, using a doubling of the best-fit chi-square as a criterion for judging the uncertainty in the parameters. The best fit value for $a$ is consistent with the inside radius of 6\,$\mu$m and outside radius of 11.5\,$\mu$m  of the field coil used in this experiment. If we assume a penetration depth of $\lambda_0=0.325\,\mu$m for Ba(Fe$_{1-x}$Co$_x$)$_2$As$_2$, then  0.28\,$\mu$m   $< \beta< 0.38\,\mu$m: $\beta \sim \lambda$.

\subsection{Comparison with a finite element calculation}
\label{sec:finite_element}

As a consistency check, we compare our results with those of a finite element calculation using the commercial software package Comsol with the ac/dc module. This module solves the equation of electromagnetism in conducting media (in S.I. units)
\begin{equation}
(i \omega \sigma -\omega^2 \epsilon_0 \epsilon_r) {\bf A} + {\rm curl}{\, \bf B}/\mu_0 \mu_r = {\bf J_e}
\label{eq:ampere}
\end{equation}
where $\sigma$ is the conductivity, $\epsilon_0$ and $\mu_0$ are the permittivity and permeability of vacuum, $\epsilon_r$ and $\mu_r$ are the relative permittivity and permeability, ${\bf B} = {\rm curl} {\, \bf A}$, and ${\bf J_e}$ is an external current source. Eq.\,(\ref{eq:ampere}) can be transformed into London's equation $\nabla^2 {\bf A}-{\bf A}/\lambda^2=0$ by setting $\sigma=-i/\mu_0 \lambda^2$ and $\omega=1$ and recognizing that the second term on the LHS of Eq.\,(\ref{eq:ampere}) is quite small. We used this procedure to solve the problem of a stripe of width $w$ and penetration depth $\lambda_b$ centered at $x=x_0$ imbedded in a bulk superconductor with penetration depth $\lambda_0$ occupying the half-space $z<0$ (with $\mu_r=1$). The field coil is modeled as a torus centered at $[0,0,z_0]$ with major radius $a$ and minor radius $b/a=0.05$. The boundary conditions were continuity of ${\bf A}$ at the internal boundaries and $\hat{n}\times{\bf A}=0$ at the external boundaries. The results of this calculation were qualitatively similar to those obtained in Ref.\,\onlinecite{kirtley2010}, although in that work the London equation $\nabla ^2 {\bf B}-\lambda^2 {\bf B}=0$ was solved, resulting in solutions that did not necessarily satisfy the condition ${\rm div} {\, \bf B}=0$. The current finite element calculations solve the London equation for the vector potential  ${\bf A}$, assuring that ${\rm div}{\,\bf B}$ = ${\rm div \, curl }{\,\bf A}=0$.
 
A plot of the resultant normalized change in susceptibility $\delta \chi/\chi$ due to a slab of superconductor with width $w/a=0.2$, $\lambda_b/a=0.1$, imbedded in a bulk superconductor with $\lambda_0/a=0.2$ is shown as the curve labelled ``Finite element" (solid symbols) in Fig.\,\ref{fig:2}. Here the parameters $z_0/a=0.17$, and $r/a=0.25$ were used. For comparison we scaled the current model predictions using $\beta^3 = (\lambda_0^2-\lambda_b^2)w=0.182a$, so that $\beta^3/\lambda_0^2 a =0.15$, (open symbols) in Fig.\,\ref{fig:2}. The finite element calculation produces a broader lineshape than the current model. This is to be expected because of the finite width assumed for the region of reduced penetration depth.  Aside from this difference the two sets of results are comparable, even though $\beta^3/\lambda_0^3=0.75$ is not much smaller than 1. These results justify, at least in retrospect, our use of the first order perturbation theory in the current work. Note that the minimum at $x_0=0$ shown in Fig. \ref{fig:2} in both the current model and in Comsol calculation has not been seen experimentally, presumably due to insufficient resolution. We thus postpone discussion of this minimum until improvements in experimental techniques make it relevant. 

 We note in this connection that in our model we keep terms of the order $\beta^3/\lambda_0^3 $ and neglect the terms ${\cal O}(\beta^6/\lambda_0^6)$ (take, e.g., Eq.\,(\ref{otvet}), use $\lambda_0$ as a unit of length to make the integral dimensionless and on the order of 1 to  see that $\psi\sim \beta^3/\lambda_0^3$).

    %%%%%%
\section{Point defect and  Green's function}
%%%%%%%%
Consider a defect as a ``vertical" line crossing the interface at a point $\bm r_0$ and extending from $z=0$ to $z=-\infty$. Physically,  such a defect affects the outside response  only from the depth on the order of the penetration depth $\lambda$. The rest of the defect line is irrelevant, so that one can consider the defect line as uniform along $z$. This is, of course, a  restriction, but it allows us to treat the  penetration depth  as two-dimensional and  to model it as 
 \begin{equation} 
\lambda^2({\bm r}) = \lambda_0^2 -  \eta^4 \delta({\bm r}-{\bm r}_0)\,,
\label{L(r)}
\end{equation}
where $\eta$ is a constant with the dimension of length. 
The solution   then provides a Green's function for a general problem of arbitrary distribution of such defects close to the sample surface relevant for the SSM technique. 
 
We have instead of Eq.\,(\ref{L3D}):
 \begin{eqnarray}
 {\bm h}_b- \lambda_0^2 {\rm \nabla}^2{\bm h}_b &=&-\eta^4\delta({\bm r}-{\bm r}_0) {\rm \nabla}^2 {\bm h}_0 \nonumber\\
 &+&
  \eta^4\delta(y-y_0)\delta^\prime(x-x_0)\hat {\bm x}\times {\rm curl} {\bm h}_0 \nonumber\\  
&+& \eta^4\delta (x-x_0)\delta^\prime (y-y_0)\hat {\bm y}\times {\rm curl} {\bm h}_0 . \qquad
\label{L3D1}
\end{eqnarray}
Here, primes denote derivatives of the delta-functions with respect to the corresponding variables and ${\bm h}_0$ is the field in absence of a defect, Eq.\,(\ref{result0}). Evaluation of the 2D FT of this equation is outlined in Appendix C: 
 \begin{eqnarray}
 {\bm h}_b^{\prime\prime} - p^2 {\bm h}_b({\bm k},z)  =- \frac{\eta^4}{\lambda_0^2}\int \frac{d{\bm q}}{4\pi^2}
e^{i({\bm q}-{\bm k}){\bm r}_0} {\bm A}\,,
\label{eq-n}
\end{eqnarray}
where $p^2=\lambda_0^{-2}+k^2$ and the vector $ {\bm A}$ is given by
\begin{eqnarray}
A_x & =& h_{0x}^{\prime\prime} -q^2h_{0x}   +(k_y-q_y)({\bm q}\times{\bm h}_0 )_z\,,\nonumber\\
A_y  &=& h_{0y}^{\prime\prime} -q^2h_{0y}   -(k_x-q_x)({\bm q}\times{\bm h}_0 )_z\,,\nonumber\\
A_z  &=& h_{0z}^{\prime\prime} -q^2h_{0z} +(q_x-k_x)   (ih_{0x}^\prime +q_xh_{0z})\nonumber\\ 
&+&(q_y-k_y)   (ih_{0y}^\prime +q_yh_{0z}) \,. 
\label{A2}
\end{eqnarray}
Since ${\bm   h}_{0} ={\bm H}e^{pz}$ with ${\bm H}$ given in Eq.\,(\ref{result0}) we have:
\begin{eqnarray}
A_{x,y} &=&2iP(P-q)q_{x,y}e^{Pz}\varphi^s({\bm q})\,, \quad P=\sqrt{\lambda_0^{-2}+q^2}\,,\quad\nonumber\\
A_z&=& 2   (P-q) \,  {\bm q}\cdot{\bm k} \,  e^{Pz}\varphi^s({\bm q}) . 
\label{A6}
\end{eqnarray}
  
 The solution of Eq.\,(\ref{eq-n}) is obtained as described in Appendix B:
 \begin{eqnarray}
   {\bm h}_b({\bm k},z)= {\bm C}e^{pz}
- \frac{ \eta^4}{\lambda_0^2}  \int  \frac{d{\bm q} }{4\pi^2} \frac{e^{Pz}}{P^2-p^2} \, {\bm A}e^{i({\bm q}-{\bm k}){\bm r}_0} .\qquad
\label{solution}
\end{eqnarray}
The conditions of div${\bm   h}_b=0$ and  of the field continuity at $z=0$   are analogous to Eqs.\,( \ref{div=0}) and ( \ref{x}) - ( \ref{z}):   
\begin{eqnarray}
 i{\bm k}\cdot{\bm C}+pC_z=
  \frac{ \eta^4}{\lambda_0^2}  \int   \frac{d{\bm q} }{4\pi^2} \frac{(i{\bm k}\cdot{\bm A}+PA_z)}{q^2-k^2} e^{i({\bm q}-{\bm k}){\bm r}_0}\,;\qquad 
\label{div=01}\\
i k_x(\varphi^s+\varphi^r) =h_{0x}+C_x-\frac{\eta^4}{\lambda_0^2}  \int   \frac{d{\bm q} }{4\pi^2}\frac{A_xe^{i({\bm q}-{\bm k}){\bm r}_0}}{q^2-k^2} \,,\qquad\label{x1}\\
  ik_y (\varphi^s+\varphi^r) =h_{0y}+C_y-\frac{ \eta^4}{\lambda_0^2}   \int   \frac{d{\bm q} }{4\pi^2} \frac{A_ye^{i({\bm q}-{\bm k}){\bm r}_0}}{q^2-k^2} \,,\qquad\label{y1}\\
  -k  (\varphi^r-\varphi^s) =h_{0z}+C_z-\frac{ \eta^4}{\lambda_0^2}  \int   \frac{d{\bm q} }{4\pi^2} \frac{A_ze^{i({\bm q}-{\bm k}){\bm r}_0}}{q^2-k^2} \,.\qquad\label{z1} 
 \end{eqnarray}
Using div${\bm   h}_0=0$ and Eq.\,(\ref{div=01}) we obtain the part of the response field due to the defect:
\begin{eqnarray}
  \psi(\bm k, {\bm  r}_0) =- \frac{2 \eta^4}{\lambda_0^2}  \int   \frac{d{\bm q} }{4\pi^2} \frac{    (P-q) \,  {\bm q}\cdot{\bm k} \,  \varphi^s({\bm q})e^{i({\bm q}-{\bm k}){\bm r}_0}}{k(p+k)(P+p)} .\qquad 
 \label{point-defect}
\end{eqnarray}

This expression can be considered as the Green's function for the general problem of a  2D  defect:
\begin{eqnarray}
G({\bm  r}, {\bm  r}_0)= \psi({\bm  r}, {\bm  r}_0)\,,\quad  G({\bm  k}, {\bm  r}_0)= \psi ({\bm  k}, {\bm  r}_0)\,.\quad 
\label{G}
\end{eqnarray}
The response potential due to a defect distributed with 
  the  area density $N({\bm r}_0) $ is
 \begin{eqnarray}
\delta\varphi^r({\bm  k}) =\int d {\bm  r}_0 N({\bm r}_0) G({\bm  k}, {\bm  r}_0) \,.\quad 
\label{G}
\end{eqnarray}

In particular, for a plane defect situated at $x_0=0$, we obtain by integrating this over $y_0$:
\begin{eqnarray}
   \delta\varphi^r =  \frac{2n \eta^4}{\lambda_0^2k(p+k)}  \int_{-\infty}^\infty   \frac{d  q_x }{2\pi } \frac{( Q-K){\bm k}\cdot{\bm Q}}{K+p} \varphi^s({\bm Q})\,.\qquad 
 \label{otvet!}
\end{eqnarray}
where $n$ is the linear  density of the point defects along the line $x=0$. This coincides with Eq.\,(\ref{otvet}) of the previous section and establishes the relation between the constants used: $\beta^3=n\eta^4$.

 Another useful example is that of a uniform slab of a width $w$ confined between the planes $x=\pm w/2$. The outside potential is obtained by integration of Eq.\,(\ref{point-defect}) over $x_0$ between $ \pm w/2$ and over $y_0$ from $-\infty$ to $\infty$:
 \begin{eqnarray}
 \delta\varphi^r = \frac{2 N \eta^4}{\pi\lambda_0^2}  \int_{-\infty}^\infty   \frac{d  q_x( Q-K){\bm k}\cdot{\bm Q} }{k(p+k)(K+p)} \,\varphi^s({\bm Q})\,
\frac{\sin\frac{(q_x-k_x)W}{2}}{ q_x-k_x } .\nonumber\\
\label{slab}
\end{eqnarray}

It is instructive to use this example to establish the relation between the  penetration depth of the ``defective" slab $\lambda_d$ and the constants we are using. To this end we write the expression (\ref{L(r)}) for a unit area of the slab cross-section represented as $N$ point defects:
\begin{eqnarray}
\lambda^2_d({\bm r}) = \lambda_0^2 -  \eta^4 \sum_\nu^N\delta({\bm r}-{\bm r}_\nu)
\label{lambda_slab}
\end{eqnarray}
where ${\bm r}_\nu$ are the positions of the point defects. Clearly,   $N=\sum \delta({\bm r}-{\bm r}_\nu)$ so that at the slab
 \begin{eqnarray}
\lambda^2_d =  \lambda_0^2 -\eta^4N\,.
\label{lambda_slab2}
\end{eqnarray}
To relate the factor $\beta$ in Eq.\,(\ref{lam(x)})  for the planar defect to  the characteristics of the slab, one takes the limit $w\to 0$ in Eq.\,(\ref{slab}) and compares the result with Eq.\,(\ref{otvet}) for planar defects to obtain:
 \begin{eqnarray}
\beta^3 = w( \lambda_0^2 -\lambda_b^2) \,.
\label{beta}
\end{eqnarray}

Similar to the slab is the case of a cylinder of a radius $R$   with penetration depth $\lambda_d$ immersed in  a material having the penetration depth $\lambda_0$:
\begin{eqnarray}
  \delta\varphi^r = \frac{  \eta^4NR}{\pi\lambda_0^2}  \int   \frac{d{\bm q}  (P-q) \,  {\bm q}\cdot{\bm k} \,  \varphi^s({\bm q})   }{k(p+k)(P+p)} \,\frac{J_1(|{\bm q}-{\bm k}|R)}{|{\bm q}-{\bm k}|}\qquad 
 \label{cylinder}
\end{eqnarray}
where $J_1$ is the Bessel function of the first order.

Another situation where the Green's function yields a straightforward solution for the outside field is a system of periodically arranged point defects. For simplicity we consider a square lattice of defects with the unit cell size $d$. The integrand in  Eq.\,(\ref{point-defect}) then contains
\begin{eqnarray}
   \sum_{{\bm r}_n} e^{i({\bm q}-{\bm k}){\bm r}_n} = \frac{4\pi^2}{d^2} \sum_{{\bm \nu} }\delta({\bm q}-{\bm k}-{\bm \nu} )\,,\qquad 
 \label{lattice}
\end{eqnarray}
where ${\bm r}_n$ are positions of the defect lattice and  ${\bm \nu}$ are reciprocal lattice vectors having $x,y$ components 
$2\pi i/d,\,\, 2\pi j/d$ with integers $i,j$ running from $-\infty$ to $\infty$ (see, e.g., Ref.\,\onlinecite{LL}).
This gives
\begin{eqnarray}
  \delta\varphi^r =- \frac{ \eta^4}{\lambda_0^2d^2k(p+k)}  \sum_{{\bm \rho}_\nu}  \frac{A_{0z} }{P+p}\Big |_{{\bm q}={\bm k}+{\bm \nu} } \,,\qquad 
 \label{otvet for lattice}
\end{eqnarray}
where $A_{0z}$ is $A_{z}$ of Eq.\,(\ref{A6}) at $z=0$.

\section{Thin films}

The problem of a linear defect in a thin film with the Pearl length $\Lambda_0=2\lambda_0^2/d$ ($d$ is the film thickness, $\lambda_0\gg d$) is formally simpler than for a bulk with a planar defect because   there is no need to consider the $z$ dependences inside the film. 

Let the superfluid density at the $y$ axis of the film at $z=0$ differ from the rest of the film. The Pearl length  than can be written as
\begin{equation} 
\Lambda(x) = \Lambda_0 -  \alpha^2 \delta(x)\,,
\label{L(x)}
\end{equation}
where the constant $\alpha$, with the dimension of length,  can be expressed in terms of the superfluid density enhancement or suppression. 

One can solve the film problem basically along the lines described in detail for the bulk case. To avoid repetitions we provide here only the  result for the magnetic potential due to the defect:  
 \begin{eqnarray}
 \psi(\bm k,x_0)=\frac{\alpha^2}{k (1+k\Lambda_0)}\int_{-\infty}^\infty \frac{dq_x }{2\pi} \frac{( \bm k\cdot\bm Q  )\varphi^s(\bm Q,x_0) }{1+\Lambda_0Q} , \qquad    
 \label{result}
\end{eqnarray}   
where $\bm Q=(q_x,k_y)$. $\varphi^s(\bm Q,x_0)$ is the 2D Fourier transform of the magnetic potential of a source at a distance $x_0$ from the defect line in the absence of a film; for a circular current source of the SQUID susceptometer, the potential is given in Eq.\,(\ref{eq:source_susc}).

\section{Discussion}

We have made an effort in this work to develop a formalism to analyze  scanning susceptometry data of superconductors containing planar defects, such as twin or grain boundaries, perpendicular to the sample surface.  
 
Superfluid density on the twin boundaries may, in some cases, be enhanced relative to the bulk.\cite{twins} Within our scheme this corresponds to the parameter $\beta>0$. In this situation, vortices should be repelled by the boundary, as observed on twinned Ba(Fe$_{1-x}$Co$_x$)$_2$As$_2$.\cite{kalisky2010,kalisky2011}  

In most cases, however, the grain boundaries attract vortices, in other words, the superconductivity is suppressed at the boundaries.  Within our scheme this is described  as $\beta <0$. The suppression of the superfluid density on grain boundaries should be observable with scanning SQUID susceptometry, but to our knowledge this experiment has not been done. 

The Green's function approach developed in Section IV may serve as a basis for studying the penetration depth of nonuniform materials, one of the outstanding problems in applying scanning susceptometry measurements to the local determination of $\lambda$. 
 
 One of the motivations for the current work is that although stripes of enhanced susceptibility associated with twin boundaries have been observed using SQUID microscopy, \cite{kalisky2010} they have not yet been seen in magnetic force microscopy.\cite{lan_and_ophir} The failure to observe stripes using MFM is puzzling, and it is hoped that the present calculations will provide guidance for future investigations.
 
\section{Acknowledgements}
We are   thankful to K. Moler for many discussions and support. We also  thank H. Bluhm for showing us how the ac/dc module in Comsol can be used for the solution of London's equations. The work  of VK  was
supported by the DOE-Office of Basic Energy Sciences, Division of Materials Sciences and Engineering under Contract No. DE- AC02-07CH11358. The work of JK was supported in part by the NSF  Grant No. PHY-0425897 and by the French NanoSciences Foundation.

%%%%%%%
\appendix
%%%%%%%

%%%%%%%%%
\section{Identities (\ref{id1}) and (\ref{id2})}
%%%%%%%%%%
The first identity is a particular case of the convolution theorem for the Fourier transform of a product:
\begin{eqnarray}
 &&FT[\delta (x)f(\bm r) ]=\int d{\bm r} e^{-i \bm {k\cdot r}}\delta (x)f(\bm r)  
 =\int_{-\infty}^\infty dy e^{-i  k_yy } f(0,y)\nonumber\\
&&= \int_{-\infty}^\infty dy e^{-i  k_yy }\int \frac{d{\bm q}}{4\pi^2} e^{i  q_y y} f({\bm q}) 
= \int_{-\infty}^\infty \frac{dq_x}{2\pi} \,f(\bm Q)\,,\label{A1}
 \end{eqnarray}
where $\bm Q=(q_x,k_y)$. Similarly, one transforms:
\begin{eqnarray}
  FT[\delta^\prime(x)  f(\bm r) ] = \int d{\bm r} e^{-i \bm {k\cdot r}}\delta^\prime (x)f(\bm r)\nonumber\\
  =- \int_{-\infty}^\infty dy e^{-i  k_yy }\frac{\partial}{\partial x}\left[e^{-i  k_xx }f(\bm r)\right]_{x=0}\nonumber\\
= i  \int_{-\infty}^\infty \frac{dq_x}{2\pi} \, (k_x-q_x)f(q_x,k_y)\,.\quad\label{A2}
 \end{eqnarray}

To Fourier transform the first term on the RHS of Eq.\,(\ref{L3Da}), we note that 
 $\hat {\bm x}\times {\rm curl} {\bm h}_0=-\hat {\bm y} \, {\rm curl}_z {\bm h}_0+\hat {\bm z} \,{\rm curl}_y {\bm h}_0$. We then have:
\begin{eqnarray}
&&   FT[\delta^\prime(x)\hat {\bm x}\times {\rm curl} {\bm h}_0 ] = 
-\hat {\bm y} \,FT[\delta^\prime(x) {\rm curl}_z {\bm h}_0] +\nonumber\\
&&\hat {\bm z} \,FT[\delta^\prime(x){\rm curl}_y {\bm h}_0] =  -i\hat {\bm y}
\int_{-\infty}^\infty \frac{dq_x}{2\pi} (k_x-q_x)(i{\bm Q}\times {\bm h}_0)_z\nonumber\\
&&+i\hat {\bm z}\int_{-\infty}^\infty \frac{dq_x}{2\pi} (k_x-q_x)[h_{0x}^\prime -iq_xh_{0z}]\,,
\label{eq3} 
 \end{eqnarray}
 where ${\bm h}_0={\bm h}_0({\bm Q},z)$ and $h_{0x}^\prime \equiv \partial {\bm h}_0 ({\bm Q},z)/\partial z$. 
 The FT of the last term on the RHS of Eq.\,(\ref{L3Da}) is:
 \begin{eqnarray}
 &&FT[\delta(x) {\rm \nabla^2}{\bm h}_0]  
  =\int_{-\infty}^\infty \frac{dq_x}{2\pi}\left[{\bm h}_0^{\prime\prime}({\bm Q} )  -Q^2 {\bm h}_0({\bm Q} )  \right].\qquad
 \label{FTb}
 \end{eqnarray}

\section{Solution of the differential Eq.\,(\ref{dif})}

The general solution of   $h^{\prime\prime}(z)-p^2h(z)=D(z)$ reads, see, e.g., Ref.\,\onlinecite{Kamke}:
\begin{eqnarray}
 && h  = C_1e^{pz}+C_2 e^{-pz}\nonumber\\
 & &+\frac{e^{pz}}{2p}\int_0^z d\zeta \,e^{-p\zeta}D(\zeta)
-\frac{e^{-pz}}{2p}\int_0^z d\zeta \,e^{p\zeta}D(\zeta) \,.\qquad
\label{A1}
\end{eqnarray}
The lower integration limits here can be chosen arbitrarily, but this choice affects the constants $C_{1,2}$ which are eventually fixed by boundary conditions. 

  Since we are dealing with a linear differential equation  (\ref{dif}) with the RHS as an integral (a sum) of the factors $e^{K\zeta}$, we can take the solution for a particular $K$ and then perform the integration (summation). Hence, we
 set $D=e^{K\zeta}$ and evaluate the integrals of Eq.\,(\ref{A1}):
\begin{eqnarray}
  h & =& e^{pz}\left[C_1-\frac{1}{2p(K-p)}\right]+ e^{-pz}\left[C_2+\frac{1}{2p(K+p)}\right]\nonumber\\
 &+&\frac{e^{Kz}}{K^2-p^2}\,.
\label{A2}
\end{eqnarray}
Since $h$ should vanish at $z\to -\infty$,   
$   C_2=- 1/2p(K+p)$\,. The solution becomes:
\begin{eqnarray}
  h =C e^{pz} +\frac{e^{Kz}}{K^2-p^2}= C e^{pz} +\frac{e^{Kz}}{q_x^2-k_x^2}\,,
\label{A4}
\end{eqnarray}
where $C$ is a redefined arbitrary constant.

%%%%%%%%%
\section{Integration over the SQUID loop}
%%%%%%%%%%

Given the FT of the response field $h_{bz}(\bm k,z_0)$, one can do the integration in real space over the area $S$ of the  circular SQUID loop:
\begin{eqnarray}
\Phi_b = \int_S   d{\bm r} h_{bz}({\bm r},z_0)=\int_S   d{\bm r}\int \frac{d{\bm k}}{4\pi^2}h_{bz}({\bm k},z_0)\,e^{i{\bm k}\cdot{\bm r}} \nonumber 
\label{Phi_s}
\end{eqnarray}
and do first the integration over the loop of a radius $r$ centered at $(x_0,0)$. To this end, one goes from the variable $ {\bm r}=(x,y)$ to ${\bm r}^\prime = (x-x_0,y)$ centered at $(x_0,0)$:  \begin{eqnarray}
  \int_S   d{\bm r} e^{i{\bm k}\cdot{\bm r}}  =e^{ik_xx_0}  \int_S   d{\bm r}^\prime e^{i{\bm k}\cdot{\bm r}^\prime}  =\frac{2\pi r}{k}\,J_1(kr) e^{ik_xx_0}.\nonumber 
\label{C1}
\end{eqnarray}
Hence,
\begin{eqnarray}
\Phi_b = \frac{r}{2\pi}\int \frac{d{\bm k}}{k}h_{bz}({\bm k},z_0)   J_1(kr) e^{ik_xx_0}.\label{C1}
\end{eqnarray}

%%%%%%%%%
\section{Fourier transform of Eq.\,(\ref{L3D1})}
%%%%%%%%%%

The LHS transforms to  
\begin{eqnarray}
 {\bm h}_b({\bm k},z)(1+\lambda_0^2k^2) -\lambda_0^2  {\bm h}_b^{\prime\prime} \,. 
\label{FT-LHS1}
\end{eqnarray}
The 2D FT of the first term at the RHS is readily shown to be
\begin{eqnarray}
&-&\eta^4\int\frac{d{\bm q}}{4\pi^2}\left[-q^2 {\bm h}_0({\bm q},z)+  {\bm h}_0^{\prime\prime} ({\bm q},z)\right]e^{i({\bm q}-{\bm k}){\bm r}_0}.\qquad 
\label{FT-RHS1}
\end{eqnarray}
The next term transforms to
\begin{eqnarray}
-\eta^4\int\frac{d{\bm q}}{4\pi^2}\Big[ {\hat {\bm y}}(q_xh_y-q_yh_x)\nonumber\\
+ {\hat {\bm z}}(ih_x^\prime + q_xh_z)\Big](q_x-k_x)e^{i({\bm q}-{\bm k}){\bm r}_0}.  
\label{FT-RHS2}
\end{eqnarray}
Here,  the arguments $({\bm q},z)$ of all Fourier components $h_i$ have been omitted for brevity along with the subscript 0 denoting unperturbed fields.
The FT of the third term on the RHS is
\begin{eqnarray}
-\eta^4\int \frac{d{\bm q}}{4\pi^2}\Big[ {\hat {\bm x}}(q_xh_y-q_yh_x)\nonumber\\
- {\hat {\bm z}}(ih_y^\prime +q_yh_z)\Big](q_y-k_y)e^{i({\bm q}-{\bm k}){\bm r}_0}.  
\label{FT-RHS3}
\end{eqnarray}

%%%%%%%%%
\section{Interaction of a vortex with a parallel defect plane in an infinite sample}
%%%%%%%%%%

An infinite vortex at ${\bm r}_v=(x_v,0)$ perpendicular to the sample surface at $z=0$ has only the $h_z(x,y)$ component, with the FT
\begin{equation}
h_{0z} =\frac{\phi_0e^{-i  k_xx_v }}{1+\lambda_0^2k^2}=\frac{\phi_0e^{-i  k_xx_v}}{\lambda_0^2p^2}\,.
\label{B1}
\end{equation}
In the presence of a planar defect at $x=0$, $\lambda$ is given in Eq.\,(\ref{lam(x)}) and the London equation  is that of (\ref{L3Da}). The total field   can be written as $ h_z = h _{0z}+  h _{bz}$ where $ h _{0z} $ is the vortex field    unperturbed by the twin boundary given in Eq.\,(\ref{B1}),  and $ h _{bz}$ is the boundary perturbation. For a weak perturbation by the boundary, we obtain:
\begin{eqnarray}
 h _b- \lambda_0^2 {\rm \nabla^2} h _b =-\beta^3\delta^\prime(x)\frac{\partial h_0}{\partial x}   -\beta^3\delta(x) {\rm \nabla^2} h_0,\qquad
\label{B2}
\end{eqnarray}
where   $ h _0 $ is  given in Eq.\,(\ref{B1}) and the subscript $z$ is omitted. After FT this gives:
\begin{eqnarray}
  h_{b }({\bm k}  )  = 
 \frac{ \phi_0\beta^3}{\lambda_0^4}  \int_{-\infty}^\infty \frac{dq_x}{2\pi} \frac{ {\bm k}\cdot{\bm Q}}{ K^2 p^2 } \,e^{-iq_xx_v}  .\qquad\qquad
\label{B3}
\end{eqnarray}

The London energy per unit length is a sum of 
 magnetic and kinetic contributions:
\begin{equation}
8\pi \varepsilon_L =\int  d{\bm r} \left[h^2+  \lambda ^2 ({\rm curl}{\bm h})^2\right].
\label{B4}
\end{equation}
Since $h=h_0+h_b$  and $\lambda^2=\lambda_0^2-\beta^3\delta(x)$, we obtain the interaction energy in linear approximation:
\begin{eqnarray}
\varepsilon_{int} &=&\int  \frac{d{\bm r}}{8\pi } \Big[2h_0h_b+ 2  \lambda_0 ^2 {\rm curl}{\bm h}_0\cdot{\rm curl}{\bm h}_b \nonumber\\
&-&\beta^3\delta(x) ({\rm curl}{\bm h}_0)^2\Big].
\label{B5}
\end{eqnarray}
%%%%%%%%
\begin{figure}[htb]
\begin{center}
\includegraphics[width=8cm]{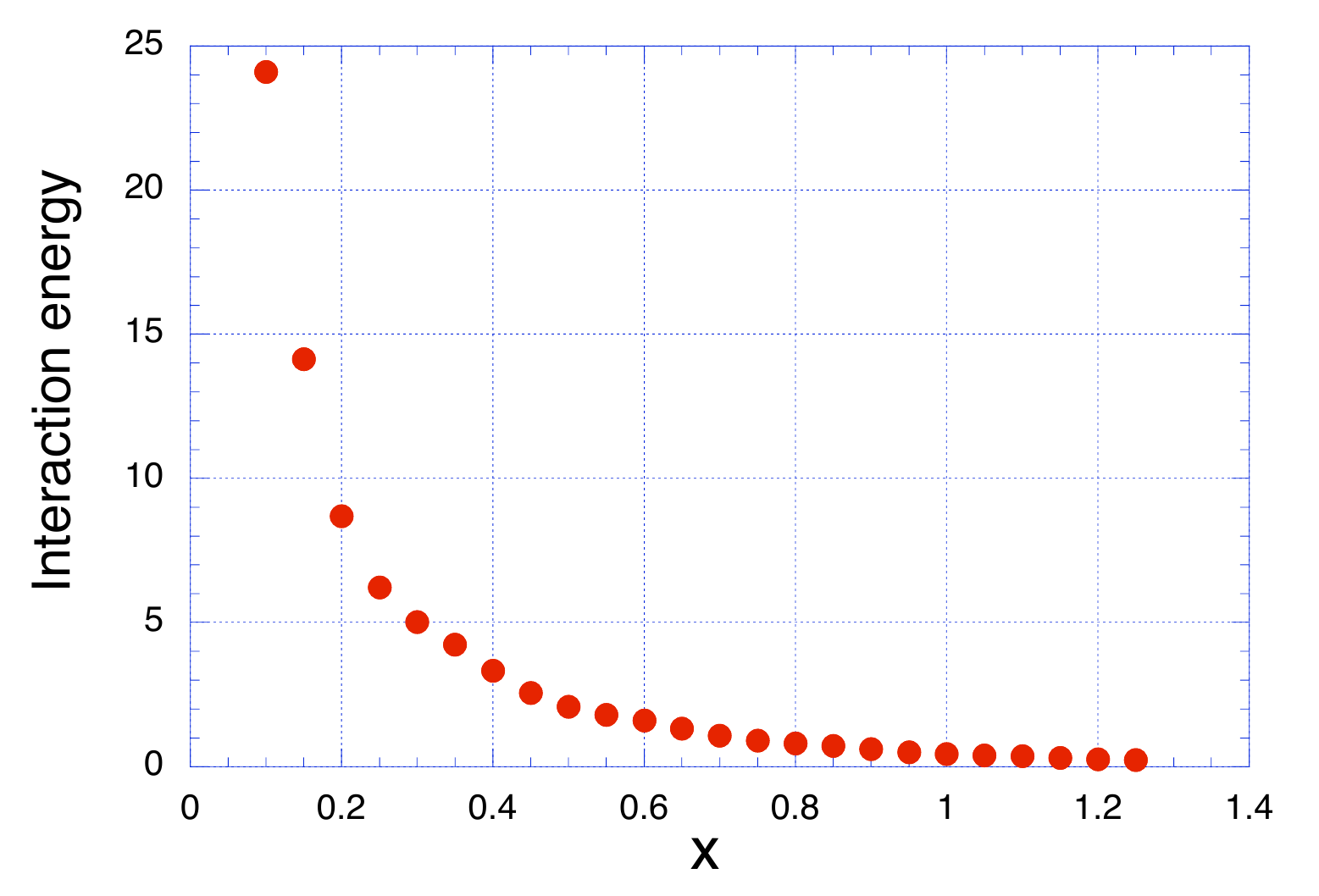}
\end{center}
\caption{(Color online)  The  interaction energy per unit length between  a vortex  parallel to a planar defect in an infinite sample  in units of $\phi_0^2\beta^3/8\pi^4\lambda_0 ^5$  as a function of $x=x_v/\lambda_0$, evaluated numerically with the help of Eq.\,(\ref{E6}) for  the superfluid density enhancement at the twin boundary.}
\label{fig3}
\end{figure}
%%%%%%%%%
We now substitute here the Fourier integrals for ${\bm h}_0$ and ${\bm h}_b$, integrate first over ${\bm r}$, and take the real part of the result:
\begin{eqnarray}
\varepsilon_{int} =\frac{\phi_0^2\beta^3}{8\pi^4\lambda_0 ^5}\int  \frac{d{\bm k}dq_x}{p^2K^2} \Big(k_xq_x\sin k_xx\,\sin q_xx  \nonumber\\
  + k_y^2 \cos k_xx\,\cos q_xx \Big). 
\label{E6}
\end{eqnarray}
Here, the integrand is dimensionless ($\lambda_0$ is used as a unit length) and  all integrals are from 0 to $\infty$ (the integrand is even in $k_x,k_y$, and $q_x$). This integral can be evaluated numerically. Fig.\,\ref{fig3} shows the resulting repulsive interaction between the vortex and  the planar defect with enhanced superfluid density, in agreement with observations reported in Ref.'s\,\onlinecite{kalisky2010,kalisky2011}.  

It is worth noting that the calculated energy $\varepsilon_{int}(x)$ diverges at $x=0$: setting $x=0$ and integrating first over $q_x$ and $k_x$  from 0 to $\infty$, one is left with  
\begin{eqnarray}
 \frac{\pi^2 }{ 4 }\int_0^m  \frac{ dk_yk_y^2}{1+k_y^2}=  \frac{\pi^2 }{ 4 }(m-\tan^{-1}m)\,,
 \label{B7}
\end{eqnarray}
which  diverges as $m\to\infty$. The divergence is an artifact of our model, which assumes that the effect of the twin plane is weak and keeps only linear terms in the correction due to the planar defect. At short distances between the vortex and the planar defect, the interaction is not weak and the model fails.

\end{document}